\documentclass[sigconf]{acmart}

\usepackage{booktabs} 
\usepackage{verbatim}
\usepackage{mdwlist}
\usepackage{soul}
\usepackage{listings}
\usepackage{color,colortbl}
\usepackage[linesnumbered,ruled,vlined]{algorithm2e}
\usepackage[english]{babel}
\usepackage[utf8]{inputenc}
\usepackage{amsmath}
\usepackage{enumitem}
\usepackage{amssymb}

\usepackage{mdwlist}
\let\emptyset\varnothing

\definecolor{dkgreen}{rgb}{0,0.6,0}
\definecolor{gray}{rgb}{0.5,0.5,0.5}
\definecolor{mauve}{rgb}{0.58,0,0.82}

\setcopyright{rightsretained}

\definecolor{amethyst}{rgb}{0.6, 0.4, 0.8}

\newcommand{\appr}{\textsc{PALOMA}\xspace}

\newcommand{\pucrs}{\raisebox{5pt}{\includegraphics[width=7pt]{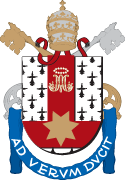}}}
\newcommand{\usc}{\raisebox{5pt}{\includegraphics[width=8pt]{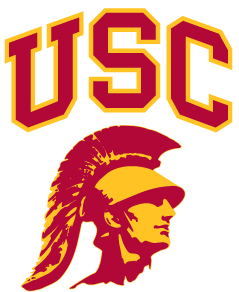}}}

\begin{document}
\title{Leveraging Program Analysis to Reduce\\User-Perceived Latency in Mobile Applications}
\copyrightyear{2018} 
\acmYear{2018} 
\setcopyright{acmcopyright}
\acmConference[ICSE '18]{ICSE '18: 40th International Conference on Software Engineering }{May 27-June 3, 2018}{Gothenburg, Sweden}
\acmPrice{15.00}
\acmDOI{10.1145/3180155.3180249}
\acmISBN{978-1-4503-5638-1/18/05}

\author{Yixue Zhao\usc, Marcelo Schmitt Laser\pucrs, Yingjun Lyu\usc, Nenad Medvidovic\usc}

\affiliation{%
\begin{tabular}{cc}
  \institution{\usc~University of Southern California} &
  \institution{\pucrs~Pontifical Catholic University of Rio Grande do Sul}\\  
  \city{Los Angeles}
  \state{CA} 
  \postcode{90089-0781}
  \country{USA} &
  \textnormal{Porto Alegre}
  \state{RS} 
  \postcode{90619-900}
  \country{Brazil} \\
  \textnormal{\{yixue.zhao, 
  yingjunl, neno\}@usc.edu} &
  \textnormal{marcelo.laser@gmail.com}
\end{tabular}
}


\begin{abstract}
Reducing  network latency in mobile applications is an effective way of improving the mobile user experience and has tangible economic benefits. This paper presents \appr, a novel client-centric technique for reducing the network latency by prefetching HTTP requests in Android apps. Our work  leverages string analysis and callback control-flow analysis to automatically instrument apps using \appr's rigorous formulation of  scenarios that address ``what'' and ``when'' to prefetch. \appr has been shown to incur significant runtime savings (several hundred milliseconds per prefetchable HTTP request), both when applied on a reusable evaluation benchmark we have developed and on real applications.
\end{abstract}


\maketitle
\vspace{-1mm}
\section{introduction}
\label{intro}
\vspace{-1mm}

In mobile computing, user-perceived latency is a critical concern as it directly impacts user experience and often has severe economic consequences. A recent report shows that a majority of  mobile users would abandon a transaction or even delete an app  if the response time of a transaction exceeds three seconds~\cite{appdynamics}. Google estimates that an additional 500ms delay per transaction would result in up to 20\% loss of traffic, while Amazon estimates that every 100ms delay would cause 1\% annual sales loss~\cite{wang2012far}.
A previous study showed that network transfer is often the performance bottleneck, and mobile apps spend  34-85\% of their time fetching data from the Internet~\cite{ravindranath2012appinsight}. A compounding factor is that mobile devices rely on \textit{wireless} networks, which can exhibit high latency, intermittent connectivity, and low bandwidth~\cite{higgins2012informed}. 

Reducing network latency thus becomes a highly effective way of improving the mobile user experience. 
In the context of mobile communication, we define \textit{latency}  as  the response time of an HTTP request. 
In this paper, we propose a novel client-centric technique for minimizing the network latency by prefetching HTTP requests in mobile apps. 
Prefetching  bypasses the performance bottleneck (in this case, network speed) and masks latency by allowing a response to a request to be generated immediately, from a local cache. 

Prefetching  has been explored in distributed systems previously. Existing approaches can be divided into four categories based on \emph{what} they prefetch and \emph{when} they do so.  \textit{(1) Server-based} techniques analyze the requests sent to the server and provide ``hints'' to the client on what to prefetch~\cite{http2push,de2007improving,padmanabhan1996using,ruamviboonsuk2017vroom}. However, most of today's mobile apps 
depend extensively on heterogeneous third-party servers. Thus, providing server-side ``hints'' is difficult, not scalable, or even impossible because app developers do not have control over the third-party servers~\cite{wang2012far}. \textit{(2) Human-based} approaches rely on  developers who have to explicitly annotate application segments that are amenable to prefetching~\cite{mickens2010crom,li2014reflection}. Such approaches are error-prone and pose significant manual burden on developers. \textit{(3) History-based} approaches build predictive models from prior requests to anticipate what request will happen next~\cite{padmanabhan1996using,behaviorcharacterization,fan1999web,Kostakos2016,WhiteSearchResultPrefetch}. Such approaches require significant time to gather  historical data. Additionally, building a precise predictive model based on history is more difficult in today's setting because the context of mobile users changes frequently. 
\textit{(4) Domain-based} approaches narrow down the problem to one specific domain. For example, approaches that focus on the social network domain~\cite{earlybird,spice:socialdrivenprefetching}  only prefetch the constant URLs in tweets based on user behavior and resource constraints. These approaches  cannot be applied to mobile apps in general.

To address these limitations of current prefetching approaches, we have developed \appr (Program Analysis for Latency Optimization of Mobile Apps),  a novel  technique that is \emph{client-centric, automatic, domain-independent, and requires no historical data}. In this paper, we focus on native Android apps because of Android's dominant market share
~\cite{androidmarketshare} and its reliance on event-driven interaction, which is the most popular style used in mobile apps today.
Our guiding insight is that an app's \emph{code} can provide a lot of useful information regarding \emph{what}  HTTP requests may occur and \emph{when}. In addition, a mobile user usually spends multiple seconds deciding what event to trigger next---a period known as ``user think time''~\cite{mickens2010crom}---providing an opportunity to prefetch HTTP requests in the background.
By analyzing an Android program, we are able to identify  HTTP requests and certain user-event sequences (e.g., \texttt{onScroll} followed by \texttt{onClick}). With that information, we can prefetch  requests that will happen next during user think time. 

User-event transitions are captured as callback control-flow~\cite{yang2015static} relationships in \appr, and we only perform very targeted, 
short-term prefetching---a single callback ahead. There are several reasons we opted for this strategy. 
First, short-term prefetching minimizes the cache-staleness problem that is commonly experienced by longer-term prefetching because the newly updated cache will be used immediately when the user transitions to the next event. Second, the information needed to send the HTTP requests (e.g., a parameter in an HTTP request that depends on user input) is more likely to be known since the prefetching occurs very close in time to the actual request. Third, short-term prefetching takes advantage of user think time \emph{between callbacks}, which has been shown to be sufficient for prefetching HTTP requests~\cite{mickens2010crom,fan1999web}. By contrast,  prefetching \emph{within the same callback} would not provide a performance gain since the relevant statements would execute within a few milliseconds of one another. 

\appr comprises four major elements. \textit{(1)~String Analysis} identifies the ``prefetchable'' requests by interpreting each URL string. \textit{(2)~Callback Analysis} determines the  prefetching points for the ``prefetchable'' requests by analyzing callback control-flow relationships. \textit{(3) App Instrumentation} modifies the original app based on the information extracted in the previous phases and outputs an optimized app that prefetches  the HTTP requests. \textit{(4) Runtime prefetching} involves the optimized app and a local proxy that is in charge of prefetching HTTP requests and managing the responses. \appr's first two elements are adaptations and extensions of existing techniques, while the latter two have been newly developed.

\appr has been evaluated for accuracy and effectiveness in two different ways. First, we developed a  microbenchmark (MBM) that isolates different prefetching conditions that may occur in an Android app. The MBM can be reused for evaluating similar future approaches. Second, we applied \appr on 32 real Android apps. Our evaluation shows that \appr exhibits perfect accuracy (in terms of precision and recall) and virtually eliminates user-perceived latency, while introducing negligible runtime overhead.

This paper makes the following  contributions:
(1) \appr, a novel client-side, automated, program analysis-based prefetching technique for mobile apps;
(2) a rigorous formulation of program analysis-based prefetching scenarios that addresses ``what'' and ``when'' to prefetch; 
(3) a comprehensive, reusable MBM to evaluate prefetching techniques for Android apps;
and (4) the implementation of an open-source, extensible framework for program analysis-based prefetching.
\appr's source code and supporting materials are publicly available~\cite{palomawebsite}.

The paper is organized as follows. Section 2 motivates the problem and defines the terms used by \appr. Sections 3 and 4 describe \appr's approach and implementation. Sections 5 and 6 detail \appr's evaluation using a benchmark and real apps. Section 7 presents related work, and Section 8 concludes the paper.
\vspace{-4mm}
\section{Background and Motivation}
\label{sec:background}
\vspace{-1mm}

In this section, we use a concrete example to introduce the fundamental building blocks and execution model of mobile apps, with a particular focus on Android. 
We then introduce our insights and motivation, followed by the definition of several key terms. 

\vspace{-2mm}
\subsection{Mobile App Example}
\label{sec:background:eg}

Mobile apps that depend on network  generally involve two key concepts: \textit{events} that interact with user inputs and \textit{network requests} that interact with remote servers. We  explain these concepts via  Listing~\ref{code:original}'s  simplified code fragment of an Android  app that responds to user interactions by retrieving weather information.

\textbf{Events:} In mobile apps, user interactions are translated to  internal app events. For instance, a screen tap  is  translated to an \texttt{onClick} event. Each event is, in turn, registered to a particular application UI object with a callback function; the callback function is executed when the event is triggered. For instance in Listing~\ref{code:original}, the button object \texttt{submitBtn} is registered with an \texttt{onClick} event (Line 9), and the corresponding callback function \texttt{onClick()} (Line 10) will be executed when a user clicks the button. Similarly, the drop-down box object \texttt{cityNameSpinner} is registered with an \texttt{onItemSelected} event that has an \texttt{onItemSelected()} callback function (Lines 5-7).


\textbf{Network Requests:} Within an event callback function, the app often has to communicate with remote servers to retrieve information. The communication is performed through network requests over the HTTP protocol in most non-realtime apps~\cite{dai2013networkprofiler}. Each HTTP request is associated with a URL field that specifies the endpoint of the request. For instance in Listing~\ref{code:original}, the \texttt{onClick} event callback sends three HTTP requests, each with a unique URL (Lines 12-14).

There are two types of URL values, depending on when the  value is known: \textit{static}  and \textit{dynamic}. For instance, \texttt{favCityId} in Listing~\ref{code:original} is static because its value is obtained statically by reading the application settings (Lines 4, 12). Similarly, \texttt{getString("domain")} reads the constant string value defined in an Android resource file~\cite{stringresource} (Line 12, 13, 14). In contrast, \texttt{cityName} is dynamic since its value depends on which item a user selects from the drop-down box \texttt{cityNameSpinner} during runtime (Lines 7, 13). Similarly, \texttt{cityId} is also a dynamic URL value (Lines 11, 14).

\begin{lstlisting}[language=Java, basicstyle=\scriptsize\ttfamily, caption={Code snippet with callbacks and HTTP requests}, label={code:original}, captionpos=b]
class MainActivity {
  String favCityId, cityName, cityId;
  protected void onCreate(){
    favCityId = readFromSetting("favCityId");//static 
    cityNameSpinner.setOnItemSelectedListener(new OnItemSelectedListener(){
      public void onItemSelected() {
      cityName = cityNameSpinner.getSelectedItem().toString();//dynamic
      }});
    submitBtn.setOnClickListener(new OnClickListener(){
      public void onClick(){
        cityId = cityIdInput.getText().toString();//dynamic
        URL url1 = new URL(getString("domain")+"weather?&cityId="+favCityId);
        URL url2 = new URL(getString("domain")+"weather?cityName="+cityName);
        URL url3 = new URL(getString("domain")+"weather?cityId="+cityId);
        URLConnection conn1 = url1.openConnection();
        Parse(conn1.getInputStream());
        URLConnection conn2 = url2.openConnection();
        Parse(conn2.getInputStream());
        URLConnection conn3 = url3.openConnection();
        Parse(conn3.getInputStream());
        startActivity(DisplayActivity.class);
    }});
  }
}
\end{lstlisting}
\vspace{-4mm}

\subsection{Motivation and Challenges}
\label{sec:background:motivation}

The motivation for \appr is that one can significantly reduce the user-perceived latency  by prefetching certain network requests. For instance, Listing~\ref{code:original} corresponds to a scenario in which a user selects a city name from the drop-down box \texttt{cityNameSpinner} (Line 7), then  clicks \texttt{submitBtn} (Line 9) to get the  city's weather information through an HTTP request. To reduce the time the user will have to wait to receive the  information from the remote server, a prefetching scheme would submit that request immediately after the user selects a city name, i.e., before the user clicks the button. 

Prefetching HTTP requests is possible for two reasons. First, an HTTP request's destination URL can sometimes be known before the actual request is sent out, such as the static URL \texttt{url1} (Line 12) in Listing~\ref{code:original}. Second, there is often sufficiently long slack between the time a request's URL value is known and when the request is sent out, due to other code's execution and the ``user think time''~\cite{mickens2010crom,fan1999web}. Prefetching in effect ``hides'' the network latency by overlapping the network requests with the slack period.

The key challenges to efficiently prefetching HTTP requests involve determining (1) which HTTP requests to prefetch, (2) what their destination URL values are, and (3) when to prefetch them. Prior work addressed these challenges by relying on various server hints~\cite{http2push,de2007improving,padmanabhan1996using}, developer annotations~\cite{mickens2010crom,li2014reflection}, and  patterns of historical user behaviors~\cite{padmanabhan1996using,behaviorcharacterization,fan1999web,Kostakos2016,WhiteSearchResultPrefetch}. Our goal is to avoid relying on such external information that may be difficult to obtain, and instead to use only program analysis on the app. 


\vspace{-2mm}
\subsection{Terminology}
\label{sec:background:term}
We define several terms needed for describing our approach to program analysis-based prefetching of network requests. 

\textbf{URL Spot} is a code statement that creates a URL object for an HTTP request based on a string denoting the endpoint of the request. Example URL Spots are Lines 12, 13, and 14 in Listing~\ref{code:original}.

\textbf{Definition Spot$_{m,n}$} is a code statement where the value of a dynamic URL string is defined, such as Lines 7 and 11 in Listing~\ref{code:original}. $m$ denotes the $m^{th}$ substring in the URL string, and $n$ denotes the $n^{th}$ definition of that substring in the code. 
For example, Line 7 would contain Definition Spot \texttt{L7$_{3,1}$} for \texttt{url2} because \texttt{cityName} is the third substring in \texttt{url2} and Line 7 is the first definition of \texttt{cityName}. A single statement of code may represent multiple Definition Spots, each of which is associated with a dynamic string used in different URLs.

\textbf{Fetch Spot} is a code statement where the HTTP request is sent to the remote server. Example Fetch Spots are Lines 16, 18, and  20. 

\textbf{Callback} is a method that is invoked \textit{implicitly} by the Android framework in response to a certain event. Example callbacks from Listing~\ref{code:original} include the \texttt{onItem\-Selected()} (Line~6) and \texttt{onClick()} (Line 10) methods. These are referred to as \textit{event handler callbacks} in Android as they respond to user interactions~\cite{androidevent}. Android also defines a set of \textit{lifecycle callbacks} that respond to the change of an app's ``life status''~\cite{androidlifecycle}, such as the \texttt{onCreate()} method at Line 3.

\begin{figure}[!b]
	\vspace{-5mm}
	\centering
	\includegraphics[width=0.285\textwidth]{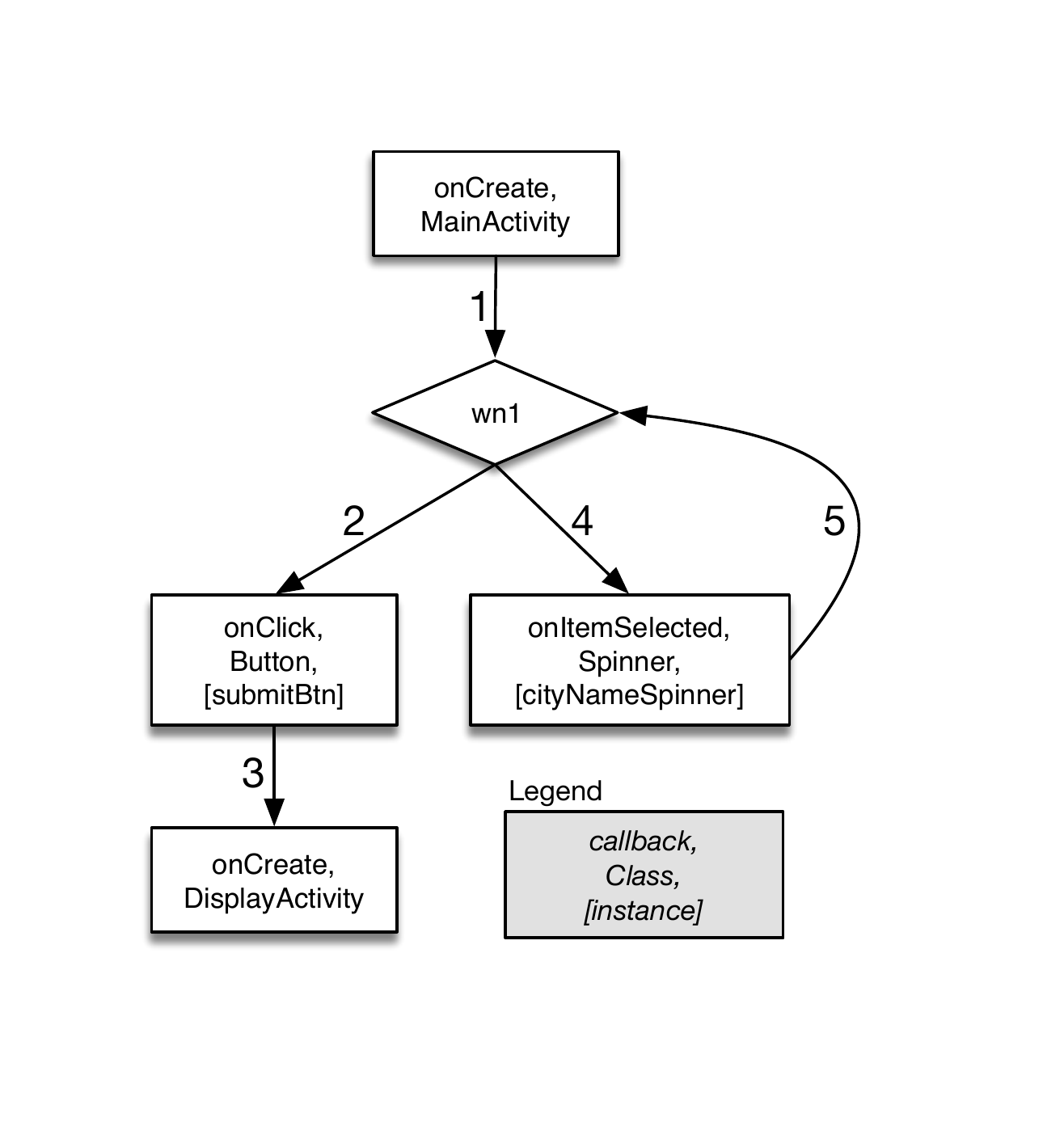}
	\vspace{-4mm}
	\caption{CCFG extracted from Listing~\ref{code:original} by GATOR~\cite{gator, yang2015static}}
	\label{ccfg}
\end{figure}

\textbf{Call Graph} is a directed graph representing the \emph{explicit} invocation relationships between procedures  in the app code.

\textbf{Target Method} is a method that contains at least one Fetch Spot. It is named that because identifying  methods that contain Fetch Spots is the target of \appr's analysis (see Section~\ref{sec:approach}). For example, the \texttt{onClick()} method is a Target Method because it contains three Fetch Spots. A Target Method may or may not be a Callback.

\textbf{Target Callback} is a Callback that can reach at least one Target Method in a Call Graph. If a Target Method itself is a Callback, it is also a Target Callback. For example, the \texttt{onClick()} Callback defined at Lines 10-22 of Listing~\ref{code:original} is a Target Callback.

\textbf{Callback Control-Flow Graph} (CCFG) represents the \emph{implicit}-invocation flow involving different Callbacks~\cite{yang2015static}. 
In a CCFG, nodes represent Callbacks, and each directed edge $f$$\rightarrow$$s$ denotes that  $s$ is the next Callback invoked  after $f$. Figure~\ref{ccfg} illustrates the CCFG extracted from Listing~\ref{code:original} using GATOR, a recently-developed analysis technique~\cite{gator, yang2015static}. A \emph{wait node} in a CCFG (e.g., \texttt{wn1} in Figure~\ref{ccfg}) indicates that the user's action is required and the event she triggers will determine which one of the subsequent callbacks is invoked. 


\textbf{Trigger Callback} is any Callback in the CCFG that is an immediate predecessor of a Target Callback with only a wait node between them. 
For instance, in Listing~\ref{code:original} the Trigger Callbacks for the Target Callback \texttt{onClick()} are \texttt{onCreate()} (path 1$\rightarrow$2) and \texttt{onItemSelected()} (path 5$\rightarrow$2). Note that \texttt{onClick()} cannot be the Trigger Callback for \texttt{DisplayActivity}'s \texttt{onCreate()} method (path 3) because there is no wait node between them.


\textbf{Trigger Point} is the program point that triggers the prefetching of one or more HTTP requests. 



\vspace{-2mm}
\section{Aproach}
\label{sec:approach}
\vspace{-1mm}

\begin{figure}[!b]
		\vspace{-4mm}
	\centering
	\includegraphics[width=0.5\textwidth]{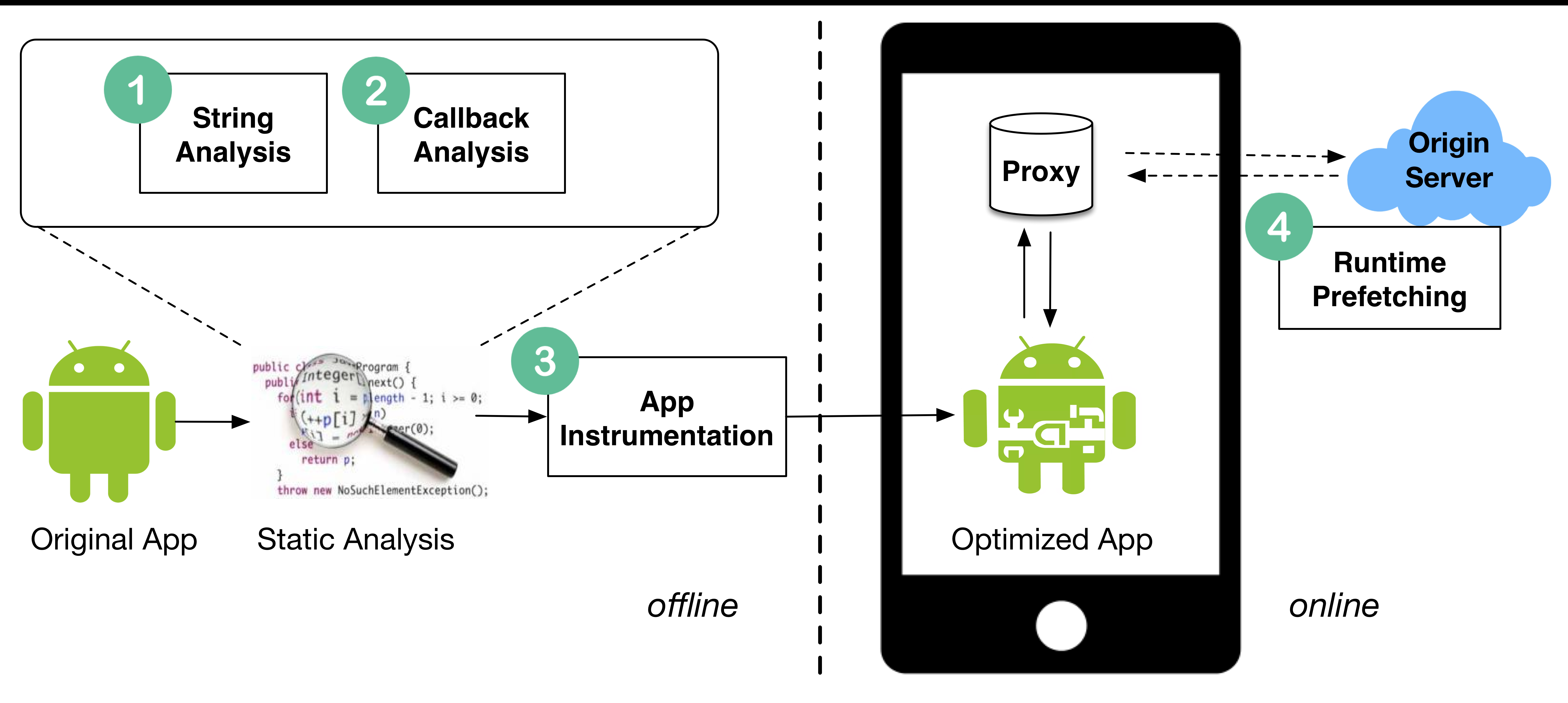}
		\vspace{-8mm}
	\caption{High-level overview of the \appr approach}
	\label{overview}
\end{figure}

This section presents \appr, a prefetching-based solution for reducing user-perceived latency in mobile apps that does not require any developer effort or remote server modifications. 
\appr is motivated by the following three challenges: (1) which HTTP requests can be prefetched, (2) what their URL values are, and (3)~when to issue prefetching requests. Our guiding insight is that 
 static program analysis can help us address all three challenges. 
 To that end, \appr employs an offline-online collaborative strategy shown in Figure~\ref{overview}. The offline component automatically transforms a mobile app into a prefetching-enabled app, while the online component issues prefetching requests through a local proxy. 

\appr has four major elements. It first performs two static analyses: it (1)~identifies HTTP requests suitable for prefetching via string analysis and (2) detects the points for issuing prefetching requests (i.e., Trigger Points) for each identified HTTP request via callback analysis. \appr then (3) instruments the app automatically based on the extracted information and produces an optimized, prefetching-enabled app. Finally at runtime, the optimized app will interact with a local proxy deployed on the mobile device. The local proxy (4) issues prefetching requests on behalf of the app and caches prefetched resources so that  future on-demand requests can be serviced immediately.
We detail these four  elements next.

\vspace{-1mm}
\vspace{-2mm}
\subsection{String Analysis}
\label{sec:approach:string}

The goal of string analysis is to identify the URL values of HTTP requests. Prefetching can only happen when the destination URL of an HTTP request is known. The key to string analysis is to differentiate between static and dynamic URL values. A static URL value is the substring in a URL whose concrete value can be determined using conventional static analysis. In contrast, a dynamic URL value is the substring in a URL whose concrete value depends on user input. For this reason, we identify the Definition Spots of dynamic URL values and postpone the actual value discovery until runtime.

\begin{figure}[t]
	\centering
	\includegraphics[width=0.52\textwidth]{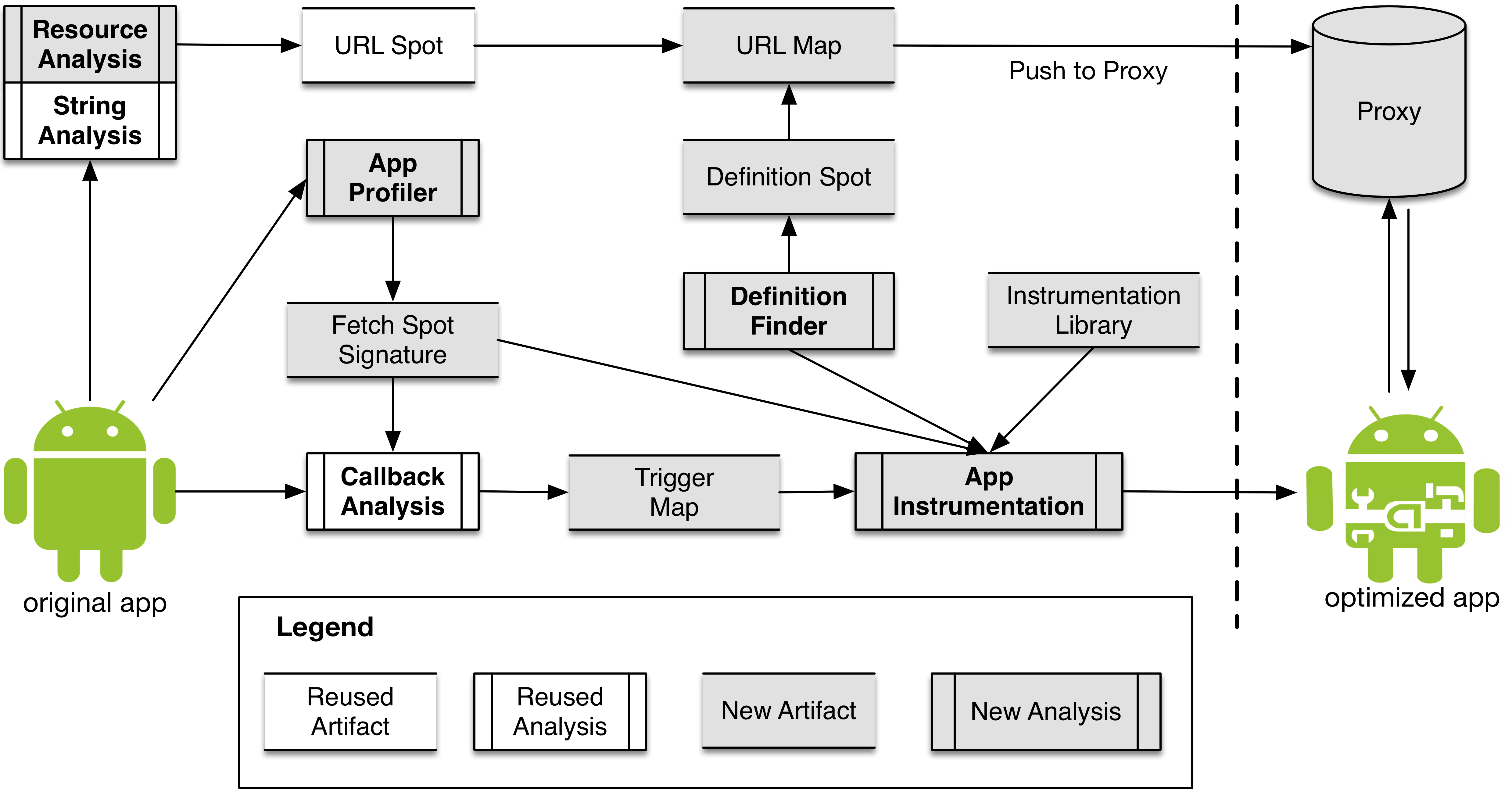}
\vspace{-6mm}
	\caption{\appr's detailed workflow. Different analysis tools employed by \appr and artifacts produced by it are depicted, with a distinction drawn between those that are extensions of prior work and newly created ones.} 
	\label{workflow}
\vspace{-5mm}
\end{figure}

As Figure~\ref{workflow} shows, the output of string analysis is a URL Map that will be used by the proxy at runtime (Section~\ref{sec:approach:prefetch}), and the Definition Spot in the URL Map will be used by the App Instrumentation step (Section~\ref{sec:approach:inst}). The URL Map relates each URL substring with its concrete value (for static values) or Definition Spots (for dynamic values). In the example of Listing~\ref{code:original}, the entry in the URL Map that is associated with \texttt{url2} would be
 
\noindent  \begin{small}\{~\texttt{url2: ["http://weatherapi/", "weather?\&cityName=", L7$_{3,1}$]}~\}\end{small}

\noindent We now explain how the URL Map is created for static and dynamic URL values.

\textbf{Static value analysis --} To interpret the concrete value of each static substring, we must find its use-definition chain and propagate the value along the chain. To do that, we leveraged a recent string analysis framework, Violist~\cite{li2015string}, that performs control- and data-flow analyses to identify the value of a string variable at any given program point.
Violist is unable to handle implicit use-definition relationships that are introduced by the Android app development framework. In particular, in Android, string values can be defined in a resource file that is persisted in the app's internal storage and retrieved during runtime. For instance in Listing~\ref{code:original}, all three URLs have a substring \texttt{getString("domain")} (Lines 12-14), which is defined in the app's resource file~\cite{stringresource}. 
\appr extends Violist to properly identify this case and include the app's resource file that is extracted by decompiling the app in the control- and data-flow analysis. In the end, the concrete value of each static substring in each URL is added to the URL Map. 




\textbf{Dynamic value analysis --} Dynamic URL values cannot be determined by static analysis. Instead, \appr identifies the locations where a dynamic value is defined, i.e., its Definition Spots. The Definition Spots are later  instrumented (see Section~\ref{sec:approach:inst})  such that the concrete values can be determined at runtime.

The key challenge in identifying the Definition Spots is that a URL string may be defined in a callback different from the callback where the URL is used. Recall that, due to the event-driven execution model, callbacks are invoked \textit{implicitly} by Android. Therefore, the control flow between callbacks on which the string analysis depends cannot be obtained by analyzing the app code statically. 
Solving the inter-callback data-flow problem is outside the scope of this paper. This is still an open problem in program analysis, because of the implicit control-flow among callbacks as well as the complex and varied types of events that can trigger callbacks at runtime, such as GUI events (e.g., clicking a button), system events (e.g., screen rotation), and background events (e.g., sensor data changes). 
Research efforts on understanding callbacks are limited to specific objectives that prevent their use for string analysis in general. Such efforts have included a focus restricted to GUI-related callbacks~\cite{yangwtg,yang2015static} (which we do use in our callback analysis, detailed in Section~\ref{sec:approach:callback}), assumption that callback control-flow can be in any arbitrary order~\cite{arzt2014flowdroid}, and analysis of the Android framework-level, but not app-level, code to construct callback summaries~\cite{Wei2017CallbackSummaries,cao2015edgeminer}.




To mitigate these shortcomings, we developed a hybrid static/dynamic approach, where the static part conservatively identifies all \emph{potential} Definition Spots,  leaving to the runtime the determination of which ones are the \emph{actual} Definition Spots. In particular, we focus on the Definition Spots of class fields because a field is a common way to pass data between callbacks. We identify all potential Definition Spots in two ways. First, if a string variable is a private member of a class, we include all the Definition Spots inside that class, such as constructor methods, setter methods, and definitions in the static block. Second, if a variable is a public member of a class, that variable can be defined outside the class and we conduct a whole-program analysis to find all assignments to the variable that propagate to the URL.

At the end of the analysis, all  substring Definition Spots for a URL are added to the URL Map. It is worth noting that although the static analysis is conservative and multiple Definition Spots may be recorded in the URL Map, the true Definition Spot will  emerge at runtime because false definitions will either be overwritten by a later true definition (i.e., a classic write-after-write dependency) or will never be encountered if they lie along unreachable paths.



\vspace{-4mm}
\subsection{Callback Analysis}
\label{sec:approach:callback}

Callback analysis  determines where  to prefetch different  HTTP requests, i.e., the Trigger Points in the app code. 
There may be multiple possible Trigger Points for a given request, depending on how far in advance the prefetching request is sent before the on-demand request is actually issued. The most aggressive strategy would be to issue an HTTP request immediately after its URL value is discovered. However, this approach 
 may lead to many redundant network transmissions: the URL value may not be used in any on-demand requests (i.e., it may be overwritten) or the callback containing the HTTP request (i.e., the Target Callback) may not be reached at runtime at all. In contrast, the most accurate strategy would be to issue the prefetching request right before the on-demand request is sent. However, this strategy would yield no  improvement in latency.

Our approach is to strike a balance between the two extremes. Specifically, \appr issues prefetching requests at the end of the callback that is the immediate predecessor of the Target Callback. Recall from Section~\ref{sec:background} that we refer to the Target Callback's immediate predecessor as Trigger Callback, because it triggers prefetching. This strategy has the dual benefit of (1) taking advantage of the ``user think time'' between two consecutive callbacks to allow prefetching to take place, while (2) providing high prefetching accuracy as the Trigger Point is reasonably close to the on-demand request.

As Figure~\ref{workflow} shows, \appr creates a Trigger Map at the end of callback analysis that is used by app instrumentation  (Section~\ref{sec:approach:inst}). The Trigger Map maps each Trigger Callback to the URLs that will be prefetched at the end of that callback. In the example of Listing~\ref{code:original}, the Trigger Map will contain two entries:

\{~[\texttt{onCreate}]: [\texttt{url1}, \texttt{url2}, \texttt{url3}]~\}

\{~[\texttt{onItemSelected}]: [\texttt{url1}, \texttt{url2}, \texttt{url3}]~\}

\noindent because both \texttt{onCreate()} and \texttt{onItemSelected()} are Trigger Callbacks that are the immediate predecessors of the Target Callback \texttt{onClick()}, which in turn contains \texttt{url1}, \texttt{url2}, and \texttt{url3}.

\begin{algorithm}[t]
\small
\DontPrintSemicolon 
\KwIn{CCFG, ECG, App}
\KwOut{TriggerMap}
$\textsc{InstrumentTimestamp}(App)$\;
$NetworkMethodLogs \gets \textsc{Profile}(App)$\;
$Signature \gets \textsc{GetFetchSignature}(NetworkMethodLogs)$\;
$Requests \gets \textsc{GetRequests}(Signature)$\;
$TriggerMap = \emptyset$\;
\ForEach{$req \in $ Requests } {
  $tarMethod \gets \textsc{GetTargetMethod}(req)$\;
  $TargetCallbacks \gets \textsc{FindEntries}(tarMethod,ECG)$\;
  \ForEach{$tarCallback \in $ TargetCallbacks}{
	  $TriggerCallbacks \gets \textsc{GetImdiatePredecessors}(tarCallback, CCFG)$\;
	  \ForEach{$trigCallback \in $ TriggerCallbacks}{
		 $TriggerMap.\textsc{Add}(trigCallback,req.url)$\;
	  }
  }
}
\Return{$TriggerMap$}\;
\caption{\sc IdentifyTriggerCallbacks}
\label{alg:callback}
\end{algorithm}
\setlength{\textfloatsep}{0pt}

Algorithm~\ref{alg:callback} details how \appr identifies Trigger Callbacks and constructs the Trigger Map. In addition to the app itself, the algorithm relies on two additional inputs, both obtained with the help of off-the-shelf-tools: the Callback Control-Flow Graph (CCFG)~\cite{yang2015static} and the Call Graph (CG)~\cite{soot}. Note that the CCFG we use in our callback analysis is restricted to GUI callbacks that are triggered by user actions (recall Section~\ref{sec:background}). However, this fits \appr's needs given its focus on user-initiated network requests. The CCFG captures the implicit-invocation flow of Callbacks in Android, and thus allows us to find the Trigger Callbacks of a given Target Callback. On the other hand, the CG, which is extracted by Soot~\cite{soot}, captures the control flow between functions, and thus allows us to locate the Callbacks that contain any given method. However, the CG does not include the direct invocations that are initiated from the Android framework. We identified such invocations from  Android's documentation and extended the CG with the resulting direct edges. An example is the \texttt{execute()}$\rightarrow$\texttt{doInBackground()} edge from the \texttt{AsyncTask} class~\cite{asynctask} that is widely used for network operations in Android. We refer to the thus extended CG as ECG.

Given these inputs, \appr first identifies the signature of a Fetch Spot, i.e., the method that issues HTTP requests, by profiling the app (Lines 1-3 of Algorithm~\ref{alg:callback}). We found that the profiling is needed because the methods that actually issue HTTP requests under different circumstances 
 can vary across  apps. For example, the \texttt{get\-InputStream()} method from Java's \texttt{URLConnection}  library may consume hundreds of milliseconds in one app, but zero  in another app where, e.g., the \texttt{getResponseCode()} method consumes several hundred milliseconds.\footnote{In this paper, we focus on  \texttt{URLConnection}, a built-in Java standard library widely used by Android developers. If the developer is using a different library and/or knows which method(s) to optimize, then  \appr's profiling step may not be needed.} Thus, we obtain the signatures by instrumenting timestamps in the app, and select the most time-consuming network operations 
 according to our profiling results. 
 Using the signatures, we then identify all  HTTP requests that the app can possibly issue (Line 4). In the example of Listing~\ref{code:original}, the signature would be \texttt{getInputStream()} and the Requests would be \texttt{conn1.getInputStream()}, \texttt{conn2.getInput\-Stream()}, and \texttt{conn3.\-getInputStream()}. We iterate through each discovered request and identify the method in which the request is actually issued, i.e., the Target Method (Line 7). Using the control flow information that the ECG provides, we  locate all possible Target Callbacks of a Target Method (Line 8). We then iterate through each Target Callback and identify all of its immediate predecessors, i.e., Trigger Callbacks, according to the CCFG (Line 10). Finally, we add each \{Trigger Callback, URL\} pair to the Trigger Map (Lines 11-12).

\vspace{-2mm}
\subsection{App Instrumentation}
\label{sec:approach:inst}

\appr instruments an app automatically based on the information extracted from the two static analyses, and produces an optimized, prefetching-enabled app as Figure~\ref{workflow} shows. At runtime, the optimized app will interact with a local proxy that is in charge of issuing prefetching requests and managing the prefetched resources~(Section~\ref{sec:approach:prefetch}).
While \appr's app instrumentation is fully automated and it does not require the source code of the app, \appr also supports app developers who have the knowledge and the source code of the app to further improve runtime latency reduction via simple prefetching hints. We describe the two instrumentation aspects next.

\vspace{3mm}
\noindent \textit{\large{3.3.1 Automated Instrumentation}}
\vspace{1mm}
\

\noindent \appr performs three types of instrumentation automatically. Each type introduces a new API that we implement in an instrumentation library. Listing~\ref{code:optimized} shows an instrumented version of the  app from Listing~\ref{code:original}, with the instrumentation code bolded. We will use this example to explain the three instrumentation tasks.

\textbf{1. Update URL Map --} This instrumentation task updates the URL Map as new values of \textit{dynamic} URLs are discovered. Recall that the values of \textit{static} URLs are fully determined and inserted into the URL Map offline. This instrumentation is achieved through a new API, \texttt{sendDefinition\allowbreak (var, url, id)}, which indicates that \texttt{var} contains the value of the $\texttt{id}^{th}$ substring in the URL named \texttt{url}. The resulting annotation is inserted right after each Definition Spot. For instance at Line 8 of Listing~\ref{code:optimized}, \appr will update the third substring in \texttt{url2} with the runtime value of \texttt{cityName}. This  ensures that the URL Map will maintain a fresh copy of each URL's value and will be updated as soon as new values are discovered.

\textbf{2. Trigger Prefetching --} This instrumentation task triggers prefetching requests at each Trigger Point. A Trigger Point in \appr is at the end of a Trigger Callback. We made this choice for two reasons: on  one hand, it makes no discernible difference in terms of performance where we prefetch within the same callback; on the other hand, placing the Trigger Point at the end is more likely to yield known URLs (e.g., when the Definition Spot is also within the Trigger Callback). \appr provides this instrumentation via the \texttt{triggerPrefetch\allowbreak (url1, ...)} API. The URLs that are to be prefetched are obtained from the Trigger Map constructed in the callback analysis (recall Section~\ref{sec:approach:callback}). For instance, \appr triggers the proxy to prefetch \texttt{url1}, \texttt{url2}, and \texttt{url3} at the end of  \texttt{onItemSelected()} (Line 9) and \texttt{onCreate()} (Line 26) of Listing~\ref{code:optimized}, which is consistent with the Trigger Map built in Section~\ref{sec:approach:callback}. 

\textbf{3. Redirect Requests --} This instrumentation task redirects all  on-demand HTTP requests to \appr's proxy instead of the origin server. This allows on-demand requests to be served from the proxy's cache, without latency-inducing network operations. The cases where the proxy's cache does not contain the response to a request are discussed in Section~\ref{sec:approach:prefetch}. The request redirection is achieved through the  \texttt{fetchFromProxy(conn)} API, where \texttt{conn} indicates the original URL connection, which is passed in case the proxy still needs to make the on-demand request to the origin server. This instrumentation replaces the original methods at each Fetch Spot: calls to the \texttt{getInputStream()} method at Lines 16, 18, and 20 of Listing~\ref{code:original} are replaced with calls to the \texttt{fetchFromProxy(conn)} method at Lines 19, 21, and 23 in Listing~\ref{code:optimized}.

\begin{lstlisting}[language=Java, basicstyle=\scriptsize\ttfamily, float=bp, caption={Example code of the optimized app}, captionpos=b, label={code:optimized}]
class MainActivity {
	String favCityId, cityName, cityId;
	protected void onCreate(){
		favCityId = readFromSetting("favCityId");//static 
		cityNameSpinner.setOnItemSelectedListener(new OnItemSelectedListener(){
			public void onItemSelected() {
			cityName = cityNameSpinner.getSelectedItem().toString();//dynamic
			(*\bfseries sendDefinition(cityName, url2, 3);*)
			(*\bfseries triggerPrefetch(url1, url2, url3);*)
			}});
		submitBtn.setOnClickListener(new OnClickListener(){
			public void onClick(){
				cityId = cityIdInput.getText().toString();//dynamic
				(*\bfseries sendDefinition(cityId, url3, 3);*)
				URL url1 = new URL(getString("domain")+"weather?&cityId="+favCityId);
				URL url2 = new URL(getString("domain")+"weather?cityName="+cityName);
				URL url3 = new URL(getString("domain")+"weather?cityId="+cityId);
				URLConnection conn1 = url1.openConnection();
				(*\bfseries Parse(fetchFromProxy(conn1));*)
				URLConnection conn2 = url2.openConnection();
				(*\bfseries Parse(fetchFromProxy(conn2));*)
				URLConnection conn3 = url3.openConnection();
				(*\bfseries Parse(fetchFromProxy(conn3));*)
				startActivity(DisplayActivity.class);
		}});
		(*\bfseries triggerPrefetch(url1, url2, url3);*)
	}
}
\end{lstlisting}

\vspace{3mm}
\noindent \textit{\large{3.3.2 Developer Hints}}
\vspace{1mm}
\

\noindent Although \appr can automatically instrument mobile apps without developer involvement, it also provides opportunities for developers to add hints in order to  better guide the prefetching. In particular, \appr enables two ways for developers to provide hints: by using its instrumentation APIs and by directly modifying its artifacts. These two approaches are described below.

\textbf{API support --} \appr's three API functions defined in the instrumentation library---\texttt{sendDefinition()}, \texttt{trigger\-Prefetch()}, and \texttt{fetchFromProxy()}---can be invoked by the developers explicitly in the app code. For instance, if a developer knows where the true Definition Spots are, she can invoke \texttt{sendDefinition()} at those locations. Developers can also invoke \texttt{triggerPrefetch()} at any program point. For example,  prefetching can happen farther ahead than is done automatically by \appr if a developer knows that the responses  to a prefetching request and its corresponding on-demand request will be identical.

\textbf{Artifact modification --} Using \appr's instrumentation APIs in the manner described above requires  modifications to the app source code. An alternative is to directly modify the artifacts generated by \appr's static analyses---Trigger Map, Fetch Spot Signature, and Definition Spot  (recall Figure~\ref{workflow})---without altering the code. 
For example, a developer can add an entry in the Trigger Map; as a result, \appr's instrumenter will automatically insert a call to \texttt{triggerPrefetch()}  at the end of the Trigger Callback specified by the developer.

We now introduce two frequently occurring instances where developers are well positioned to provide prefetching hints with very little manual effort. These hints can be provided using either of the above two approaches.





\textbf{Prefetching at app launch --} Launching an app  may take several seconds or more because many apps request remote resources, typically toward the end of the launch process. The URLs of the launch-time requests are usually statically known, but the  ways in which  the URL values can be obtained are highly app-dependent. For instance,  apps may retrieve the request URLs from a configuration file or a local database. Supporting  those cases in \appr's string analysis would mean that \appr must understand the semantics of each individual app, which is not a reasonable requirement. However, a practical alternative is for developers to provide prefetching hints because they understand their own apps' behavior. One way  developers could implement this is to insert into the URL Map additional static URLs and then call \texttt{triggerPrefetch()} at the beginning of \texttt{onCreate()}, which for \appr's purposes can be treated as the  app entry point in most Android applications. 

\textbf{Prefetching for ListView --} 
The \texttt{ListView} class~\cite{listview} is commonly used in Android apps to display the information of a list of items. The app ``jumps'' to another page to display further information based on the item a user selects in the list. The URL fetched for the page to which the app ``jumps'' is typically only known after the user selects the item in the list. Ordinarily, this would prevent prefetching. However, Android apps tend to exhibit two helpful trends. First, the list view usually displays similar types of information. Second, the further information obtained by selecting an item is related to the information displayed in the list itself. Based on these observations, we identified and are exploiting in \appr similar patterns in the URLs for the list and the subsequent page. Consider a wallpaper app for illustration: The URL that is fetched to render an item in the list view may be ``image1Url\_small.jpg'', while the URL that is fetched after the user selects \texttt{image1} may be ``image1Url\_large.jpg''. Based on this pattern, we have explored manually adding Definition Spots of the URLs that are fetched in the list view and sending modified values to the proxy, such as replacing ``small'' with ``large'' in the wallpaper example. 

\vspace{-3mm}
\subsection{Runtime Prefetching}
\label{sec:approach:prefetch}

\appr's first three phases are performed offline. By contrast, this phase captures the interplay between the optimized apps and \appr's proxy to prefetch the HTTP requests at runtime. The instrumented methods in an optimized app trigger the proxy to perform corresponding functions. We now use the example from Listing~\ref{code:optimized} to show how the three instrumented functions from Section~{3.3.1} interact with the proxy.

\textbf{1. Update URL Map -- } When the concrete value of the \emph{dynamic} URL is obtained at runtime, the inserted instrumentation method  \texttt{sendDefinition\allowbreak (var, url, id)}  is executed and the concrete runtime value is sent to the proxy. In response, the proxy  updates the corresponding URL value in the URL Map. For instance in Listing~\ref{code:optimized}, when a user selects a city name from the \texttt{cityNameSpinner} (Line 7), the concrete value of \texttt{cityName} will be known, e.g., ``Gothenburg''. Then \texttt{cityName} is sent to the proxy (Line 8)  and the URL Map entry for \texttt{url2}  will be updated to \{~\texttt{url2: ["http://weatherapi/", "weather?\&cityName=",\allowbreak \textbf{"Gothenburg"}]}~\}. 

\textbf{2. Trigger Prefetching -- } When the inserted instrumentation method \texttt{trigger\allowbreak Prefetch(url1,...)}  is executed, it triggers the proxy to perform \textsc{TriggerPrefetch} as shown in Algorithm~\ref{alg:trigger}. For each request that is sent to the proxy by \texttt{trigger\-Prefetch\allowbreak(url1,...)}, the proxy  checks if the whole URL of the request is known but the response to the request has not yet been cached (Line 2). If both conditions are met, a ``wait'' flag is set  in the cache for that request  (Line 3). This ensures that duplicated requests will not be issued in the case when the on-demand request is made by the user before the response to the prefetching request has been returned from the origin server. 
In the example of Listing~\ref{code:optimized}, when the app reaches the end of \texttt{onCreate} (Line 26), it triggers the proxy to perform \texttt{TriggerPrefetch(url1,url2,url3)}. Only \texttt{url1} meets both conditions at Line 2 of Algorithm~\ref{alg:trigger}:  the URL value is concrete (it is, in fact, a \emph{static} value) and the response is not in the cache. The proxy thus sets the ``wait'' flag for \texttt{url1} in the cache, prefetches \texttt{url1} from the origin server, stores the response in the cache, and finally sends an ``unwait'' signal to the on-demand request that is waiting for the prefetched request (Line 3-6). Thereafter, when the user selects a city name from the dropdown box, \texttt{onItemSelected} (Line 6 of Listing~\ref{code:optimized}) will be triggered. At the end of \texttt{onItemSelected} (Line 9),  \texttt{TriggerPrefetch(url1,url2,url3)} is invoked again and \texttt{url2} will be prefetched because its URL is known (its \textit{dynamic} value obtained at Line 8) and has not been previously prefetched. In contrast, the value of \texttt{url1} is known at this point but \texttt{url1} was already prefetched at Line 26, so the proxy will not prefetch \texttt{url1}.

\begin{algorithm}[t]
\small
\DontPrintSemicolon 
\KwIn{Requests}
\ForEach{$req \in $ Requests } {
		\If{$\textsc{IsKnown}(req.url)$ \textbf{and} $\neg$$\textsc{IsCached}(req)$}{
			$\textsc{SetWaitFlag}(req)$\;
			$response \gets req.\textsc{FetchRemoteResponse}()$\;
			$cache.\textsc{Put}(req,response)$\;
			$\textsc{UnWait}(req)$\;
		}
	} 
\caption{\sc TriggerPrefetch}
\label{alg:trigger}
\end{algorithm}

\textbf{3. Redirect Requests -- } When the on-demand request is sent at the Fetch Spot, the replaced function \texttt{fetchFromProxy(conn)} will be executed, and it will in turn trigger the proxy to perform \textsc{ReplacedFetch} as shown in Algorithm~\ref{alg:fetch}. If the request has a corresponding response in the cache, the proxy will first check the ``wait'' flag for the request. If the flag is set, the proxy will wait for the signal  of the prefetching request (Line 3) and will return the response of the prefetching request when it is back from the origin server (Line 4). If the ``wait'' flag has not been set, the response is already in the cache and the proxy returns the response immediately with no network operations involved (Line 4). Otherwise, if the cache does not contain the response to the request, the proxy issues an on-demand request using the original URL connection \texttt{conn} to fetch the response from the origin server, stores the response in the cache, and returns the response to the app (Line 6-8). 
For instance in Listing~\ref{code:optimized}, if a user clicks \texttt{submitBtn}, \texttt{fetch\allowbreak FromProxy(conn)} will be executed to send on-demand requests for \texttt{url1}, \texttt{url2}, \texttt{url3}  to the proxy (Lines 19, 21, and 23 of Listing~\ref{code:optimized}). The proxy in turn returns the responses to \texttt{url1} and \texttt{url2} from the local cache immediately because \texttt{url1} and \texttt{url2} are prefetched at Lines 26 and 9 respectively, as discussed above. \texttt{url3} is not known at any of the Trigger Points, so the response to \texttt{url3} will be fetched from the origin server on demand as in the original app. Note that if a user did not select a city name from the dropdown box before clicking \texttt{submitBtn}, \texttt{onItemSelected} will not be triggered, meaning that Lines 8 and 9 of Listing~\ref{code:optimized} will not be executed. In this case, only the response for \texttt{url1} will be returned from the cache (prefetched at Line 26) while the on-demand requests for \texttt{url2} and \texttt{url3} will be routed to the origin server.

\begin{algorithm}[t]
\small
\DontPrintSemicolon 
\KwIn{$req \in Requests$}
\KwOut{$response \in Responses$}
	\If{$\textsc{IsCached}(req)$}{
		\If{$\textsc{GetWaitFlag}(req)$ \textbf{is} $TRUE$}{
				$\textsc{Wait}(req)$\;
		}
		\Return{$cache.\textsc{GetResponse}(req$)}
	}
	\Else{
		$response \gets req.\textsc{FetchRemoteResponse}()$\;
		$cache.\textsc{Put}(req,response)$\;
		\Return{$response$}\;
	}
\caption{\sc ReplacedFetch}
\label{alg:fetch}
\end{algorithm}

\vspace{-1mm}
\section{Implementation}
\label{sec:implementation}
\vspace{-1mm}

\appr has been implemented by reusing and extending several off-the-shelf tools, and integrating them with newly implemented functionality. PALOMA's string analysis extends the string analysis framework Violist~\cite{li2015string}. The callback analysis is implemented on top of the program analysis toolkit GATOR~\cite{gator}, and by extending GATOR's CCFG analysis~\cite{yang2015static}. \appr's  instrumentation component is a stand-alone Java program that uses Soot~\cite{soot} to instrument an app. The proxy is built on top of the Xposed framework~\cite{xposed} that provides mechanisms to ``hook'' method calls. The proxy intercepts the methods that are defined in \appr's instrumentation library and replaces their bodies with corresponding methods implemented in the proxy. The total amount of newly added code to extend existing tools, implement the new functionality, and integrate them together in \appr is 3,000 Java SLOC.
\vspace{-1mm}
\section{Microbenchmark Evaluation}
\label{sec:benchmark_eval}
\vspace{-1mm}

In this section, we describe the design of a microbenchmark (MBM) containing a set of test cases, which we used to evaluate \appr's \textit{accuracy} and \textit{effectiveness}. 

MBM thoroughly covers the space of prefetching options, wherein each test case contains a single HTTP request and differs in whether and how that request is prefetched. The MBM is restricted to individual HTTP requests because the requests are issued and processed independently of one another. This means that \appr will process multiple HTTP requests simply as a sequence of individual requests; any concurrency in their processing that may be imposed by the network library and/or the OS is outside \appr's purview. In practice, the look-up time for multiple requests  varies slightly from one  execution of a given app to the next. However, as shown in Section~\ref{sec:benchmark_eval:result}, the look-up time required by \appr would not be noticeable to a user even with a large number of requests. As we will show in Section~\ref{sec:realapp_eval}, the number of HTTP requests in real apps is typically bounded. Moreover, \appr only maintains a small  cache that is emptied every time a user quits the app. 

In the rest of this section, we will first lay out the  goals underlying the MBM's design (Section~\ref{sec:benchmark_eval:principle}), and then present the MBM (Section~\ref{sec:benchmark_eval:test_cases}). Our evaluation results show that \appr achieves perfect accuracy when applied on the MBM, and leads to significant latency reduction with negligible runtime overhead (Section~\ref{sec:benchmark_eval:result}).


\vspace{-2mm}
\subsection{Microbenchmark Design Goals}
\label{sec:benchmark_eval:principle}

The MBM is designed to evaluate two fundamental aspects of \appr: \textit{accuracy} and \textit{effectiveness}. 

\appr's \textbf{accuracy} pertains to the relationship between \textit{pre\-fetchable} and actually
\textit{prefetched} requests. Prefetchable requests are requests whose URL values are known before the Trigger Point 
and thus can be prefetched. Quantitatively, we capture accuracy via the dual measures of \emph{precision} and \emph{recall}. 
Precision indicates how many of the  requests that \appr tries to prefetch at a given Trigger Point were actually prefetchable.
On the other hand, recall indicates how many requests are actually prefetched by \appr out of all the prefetchable requests at a given Trigger Point.


\appr's \textbf{effectiveness} is also captured by two measures: the runtime overhead introduced by \appr and the latency reduction achieved by it. Our objective is to minimize the runtime overhead while maximizing the reduction in user-perceived latency. 


\vspace{-2mm}
\subsection{Microbenchmark Design}
\label{sec:benchmark_eval:test_cases}

The MBM is built around a key concept---\emph{prefetchable}---a request whose whole URL is known before a given Trigger Point. We refer to the  case where the request is prefetchable and the response is used by the app as a \emph{hit}. Alternatively, a request may be prefetchable but the response is not used because the definition of the URL is changed after the Trigger Point. We call this a \emph{non-hit}.  The 
 MBM aims to cover all  possible cases of  \emph{prefetchable}  and \emph{non-prefetchable} requests, including \emph{hit} and \emph{non-hit}.


There are three factors that affect whether an HTTP request is prefetchable: (1) the number of dynamic values in a URL; (2) the number of Definition Spots for each dynamic values; and (3) the location of each Definition Spot relative to the Trigger Point. We now formally define the properties of \emph{prefetchable} and \emph{hit} considering the three factors. The formal definitions will let us succinctly describe test cases later.

\textbf{Formal Definition.} Let $M$ be the set of Definition Spots before the Trigger Point and $N$  the set of Definition Spots after the Trigger Point, which is within the Target Callback (recall Sections~\ref{sec:background} and \ref{sec:approach:callback}). Let us assume that a URL has $k\geq 1$ dynamic values. (The case where $k = 0$, i.e., the whole URL is static, is considered separately.) Furthermore, let us assume that the dynamic values are the first $k$  values in the URL.\footnote{This assumption is used only to simplify our formalization. The order of the values in a URL has no impact on whether the URL is prefetchable and can thus be arbitrary.} The $i^{th}$ dynamic value ($1 \leq i \leq k$) has $d_i\geq 1$ Definition Spots in the whole program. A request is
\begin{itemize}
\item \underline{\textit{prefetchable}} ~~iff ~~$\forall$$i$~$\exists$($j\in[1..d_i]$)~$\mid$~DefSpot$_{i,j}\in M$\\
(every dynamic value has a DefSpot before Trigger Point) 
\item \underline{\textit{hit}} ~~iff~~ \emph{prefetchable} $\land$ $\forall$($j\in[2..d_i]$)~$\mid$~DefSpot$_{i,j}\in M$\\
(all dynamic value DefSpots are before Trigger Point)

\item \underline{\textit{non-hit}} ~~iff~~ \emph{prefetchable} $\land$ $\exists$($j\in[2..d_i]$)~$\mid$~DefSpot$_{i,j}\in N$\\
(some dynamic value DefSpots are after Trigger Point)


\item \underline{\textit{non-prefetchable}} ~~iff~~ 
$\forall$($j\in[1..d_i]$)~$\exists$$i$~$\mid$~DefSpot$_{i,j}\in N$\\
(all DefSpots for a dynamic value are after Trigger Point)
\end{itemize}

Without loss of generality, MBM covers all cases where $k \leq 2$ and $d_i \leq 2$. We do not consider cases where $k > 2$ or $d_i > 2$ because we only need two dynamic values to cover the \emph{non-prefetchable} case---where some dynamic values are unknown at the Trigger Point---and two Definition Spots to cover the \emph{non-hit} case---where some dynamic values are redefined after the Trigger Point.

\begin{figure}[!t]
	\centering
	\includegraphics[width=0.9\columnwidth]{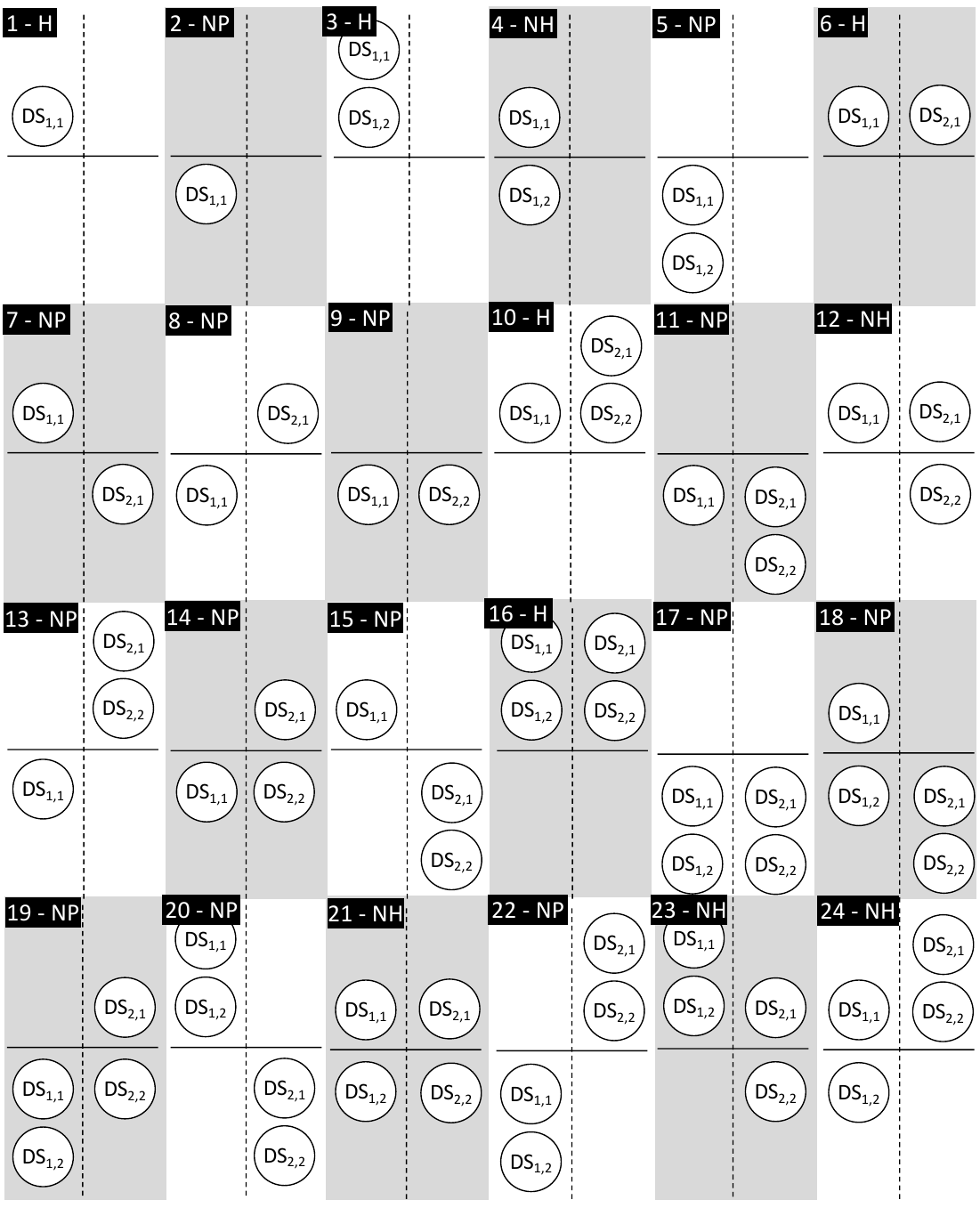}
		\vspace{-4mm}
	\caption{The 24 test cases covering all configurations involving dynamic values. The horizontal divider denotes the Trigger Point, while the vertical divider delimits the two dynamic values. The circles labeled with ``DS$_{i,j}$'' are the locations of the Definition Spots with respect to the Trigger Point. ``H'' denotes a \emph{hit}, ``NH'' denotes a \emph{non-hit}, and ``NP'' denotes a \emph{non-prefetchable} request.}
	\label{fig:MBM}
\end{figure}

There are a total of 25 possible cases involving configurations with $k \leq 2$ and $d_i \leq 2$. The  simplest case is when the entire URL is known statically; we refer to it as case 0. The remaining 24 cases are diagrammatically encoded in Figure~\ref{fig:MBM}: the two dynamic URL values  are depicted with circles and delimited with the vertical line; the location of the Trigger Point is denoted with the horizontal line; and the placement of the circles  marks the locations of the dynamic values' Definition Spots (``DS$_{i,j}$'' in the figure) with respect to the Trigger Point.  These 24 cases can be grouped as follows:
\vspace{-1mm}
\begin{itemize}
\item single dynamic value -- cases 1-5;
\item two dynamic values, one Definition Spot each -- cases 6-9;
\item two dynamic values, one with a single Definition Spot, the other with two -- cases 10-15; and
\item two dynamic values, two definition spots each -- cases 16-24.
\end{itemize}
\vspace{-1mm}
Each case is labeled with its \emph{prefetchable/hit} property (``H'' for \emph{hit}, ``NH'' for \emph{non-hit}, and ``NP'' for \emph{non-prefetchable}). Of particular interest are the six cases---0, 1, 3, 6, 10, and 16---that represent the  \emph{hit}s that should allow \appr to prefetch the corresponding requests and hence significantly reduce the user-perceived latency.

\vspace{-2mm}
\subsection{Results}
\label{sec:benchmark_eval:result}

We implemented the MBM as a set of Android apps along with the remote server to test each of the 25 cases. The server is built with Node.js and  deployed on the Heroku cloud platform~\cite{heroku}. The apps interact with the server to request information from a dataset in MongoDB~\cite{mongodb}. The evaluation was performed  on the 4G network.  
The testing device was Google Nexus 5X running Android 6.0. Overall, our evaluation showed that \appr achieves 100\% precision and recall without exception, introduces negligible overhead, and can reduce the latency to nearly zero under appropriate conditions (the \emph{hit} cases discussed above).

\begin{table}[t]
  \caption{Results of \appr's  evaluation using MBM apps covering the 25 cases discussed in Section~\ref{sec:benchmark_eval:test_cases}.  ``SD'', ``TP'', and ``FFP''  denote the runtimes of the three \appr instrumentation methods. 
   ``Orig''  is the time required to run the original app.  ``Red/OH''  represents the reduction/overhead in execution time when applying \appr. } 
  \vspace{-3mm}
  	\label{tbl:benchmark}
  	\centering
 	\resizebox{.98\linewidth}{!}{
\begin{tabular}{|c|c|c|c|c|c|}
\hline
\textbf{Case}              & \textbf{SD (ms)}     & \textbf{TP (ms)} & \textbf{FFP (ms)}        & \textbf{Orig (ms)}            & \textbf{Red/OH}       \\ \hline
\rowcolor[HTML]{C0C0C0} 
0                         & N/A                      & 2            & \textbf{1}                & 1318                                                 & 99.78\%                      \\ \hline
\rowcolor[HTML]{C0C0C0} 
{\color[HTML]{000000} 1}  & {\color[HTML]{000000} 0} & {\color[HTML]{000000} 5}     &{\color[HTML]{000000} \textbf{0}} & {\color[HTML]{000000} 15495} &  {\color[HTML]{000000} 99.97\%} \\ \hline
2                          & 0                        & 1                & 2212            & 2659                                                       &            16.81\%                  \\ \hline
\rowcolor[HTML]{C0C0C0} 
3                         & 1                      & 4                & \textbf{1}                 & 781                                             & 99.24\%                      \\ \hline
4                          & 2                        & 5                   & 611                & 562                                                  &         -9.96\%                     \\ \hline
5                          & 0                        & 2                  & 2588               & 2697                                                  &          3.97\%                    \\ \hline
\rowcolor[HTML]{C0C0C0} 
6                         & 1                        & 4                      & \textbf{2}            & 661                                            & 98.95\%                      \\ \hline
7                          & 1                        & 4                      & 2237                   & 2399                                          &         6.54\%                     \\ \hline
8                          & 1                        & 9                         & 585              & 568                                              &               4.75\%               \\ \hline
9                          & 2                        & 2                        & 611                      & 584                                       &       -5.31\%                       \\ \hline
\rowcolor[HTML]{C0C0C0} 
{\color[HTML]{000000} 10} & {\color[HTML]{000000} 1} & {\color[HTML]{000000} 5}  & {\color[HTML]{000000} \textbf{0}}   & {\color[HTML]{000000} 592}    & {\color[HTML]{000000} 98.99\%} \\ \hline
11                         & 2                        & 2                & 2813                 & 2668                                                  &         -5.58\%                     \\ \hline
12                         & 2                        & 6                    & 546                 & 610                                                &               8.16\%               \\ \hline
13                         & 2                        & 3                    & 2478               & 2753                                                &            10.87\%                  \\ \hline
14                         & 3                        & 3                    & 549                & 698                                                 &            20.49\%                  \\ \hline
15                         & 5                        & 1                    & 631                  & 570                                               &           -11.75\%                   \\ \hline
\rowcolor[HTML]{C0C0C0} 
16                        & 1                        & 11                        & \textbf{0}                & 8989                                    & 99.87\%                        \\ \hline
17                         & 0                        & 3                  & 418                        & 555                                           &      31.83\%                        \\ \hline
18                         & 2                        & 6                 & 617                        & 596                                            &      -4.87\%                        \\ \hline
19                         & 4                        & 6                     & 657                      & 603                                          &     -10.61\%                         \\ \hline
20                         & 1                        & 3                       & 620                       & 731                                       &          17.15\%                    \\ \hline
21                         & 2                        & 10                          & 611              & 585                                            &       -6.50\%                       \\ \hline
22                         & 2                        & 7                     & 737                   & 967                                             &      29.62\%                        \\ \hline
23                         & 2                        & 9                      & 608                      & 607                                         &      -1.98\%                        \\ \hline
24                         & 1                        & 10                        & 611                 & 715                                           &    14.95\%                          \\ \hline
\end{tabular}
 	}
\end{table}

Table~\ref{tbl:benchmark} shows the  results of each test case corresponding to Figure~\ref{fig:MBM}, as well as case 0 in which the entire URL value is known statically.  Each execution value is the average of multiple executions of the corresponding MBM apps. The highlighted test cases are the \emph{hit} cases that should lead to a significant latency reduction. The columns ``SD'', ``TP'', and ``FFP'' show the average times spent in the corresponding \appr instrumentation  methods   in the optimized apps---\texttt{sendDefinition()}, \texttt{trigger\-Prefetch()}, and \texttt{fetchFrom\-Proxy()}, respectively (recall Section~\ref{sec:approach:inst}). 
 The ``Orig'' column shows the execution time of the method invoked at the Fetch Spot in the original app, such as \texttt{getInputStream()}.  

The final column in the table, labeled ``Red/OH'' shows the percentage reduction in execution time when \appr is applied on each MBM app. The reduction is massive in each of the six \emph{hit} cases ($\geq$99\%). It was interesting to observe that applying \appr also resulted in reduced average execution times in 11 of the 19 \emph{non-hit} and \emph{non-prefetchable} cases. The more expected scenario occurred in the remaining eight of the \emph{non-hit} and \emph{non-prefetchable} cases: applying \appr introduced an execution overhead (shown as negative values in the table). The largest runtime overhead introduced by \appr was 149ms in case 11, where the original response time was 2,668ms. This value was due to a couple of  outliers in computing the average execution time, and it may be attributable to factors in our evaluation environment other than \appr, such as network speed; the remaining measurements were significantly lower.  However, even this value is actually not prohibitively expensive: recall that \appr is intended to be applied in cases in which a  user already typically spends multiple seconds deciding what event to trigger next~\cite{mickens2010crom}. 

\vspace{-1mm}
\section{Third-Party App Evaluation}
\label{sec:realapp_eval}
\vspace{-1mm}

We also evaluated \appr on third-party Android apps to observe its behavior in a real-world setting.
We used the same execution setup as in the case of the MBM. 
We selected 32 apps from the Google Play store~\cite{googleplay}.  
We made sure  that the selected apps span a range of application categories---Beauty, Books \& Reference, Education, Entertainment, Finance,  Food \& Drink, House \& Home, Maps \& Navigation, Tools, Weather, News \& Magazines, and Lifestyle---and vary in sizes---between 312KB and 17.8MB. 
The only other constraints in selecting the apps were that they  were executable, relied on the Internet, and could be processed by Soot.\footnote{Soot is occasionally unable to process an Android app for reasons that we were unable to determine. This issue was also noted by others previously.}      

We asked two Android users to actively use the 32 subject apps for two minutes each, and recorded the resulting usage traces.\footnote{While the average app session length varies by user and app type (e.g.,~\cite{statista}), two minutes was sufficiently long to observe representative behavior and, if necessary, to extrapolate our data to longer sessions.} We then  re-ran the same traces on the apps  multiple times, to account for  variations caused by the runtime environment. Then we instrumented the apps  using \appr and repeated the same steps the same number of times. Each session started with app (re)installation and exposed all app options to users. As in the case of the MBM, we  measured and  compared the response times of the methods at the Fetch Spots between the original  and  optimized apps.

Unlike in the case of the MBM, we do not have the ground-truth data for the third-party apps. Specifically, the knowable URLs   at the Trigger Points would have to be determined manually, which is prohibitively time-consuming and error prone. In fact, this would boil down to manually performing inter-callback data-flow analysis (recall Section~\ref{sec:approach:string}). For this reason, we measured only two aspects of applying \appr on the third-party apps: the \emph{hit rate} (i.e., the percentage of requests that have been \emph{hit} out of all triggered requests) and the resulting \emph{latency reduction}. Table 2 depicts the averages, outliers (min and max values), as well as the standard deviations obtained across all of the runs of the 32 apps.

Overall, the results show that \appr achieves a significant latency reduction with a reasonable hit rate. There are several  interesting outlier cases. The minimum hit-rate  is only 7.7\%. The reason is that the app in question fetches a large number of  ads at runtime whose URLs are non-deterministic, and only a single static URL is prefetched outside those. There are  four additional apps whose hit rate is below 20\% because those apps are list-view apps, such as a wallpaper app (recall Section~\ref{sec:approach:inst}), and they fetch large numbers of requests at the same time. In \appr, we set the threshold for the maximum number of requests to prefetch at once to be 5. This parameter can be increased, but that may impact device energy consumption, cellular data usage, etc. This is a trade-off that will require further study. 

Similarly to the MBM evaluation, \appr achieves a reduction in latency of nearly 99\% on average for ``hit'' cases. Given the average execution time for processing a single request across the 32 unoptimized apps of slightly over 800ms, prefetching the average of 13.28 requests at runtime would reduce the total app execution time by nearly 11s, or 9\% of a two-minute session. Note that the lowest latency reduction was 87.41\%. This was caused by on-demand requests that happen before the prefetching request is returned (recall the discussion in Section~\ref{sec:approach:prefetch}). In those cases, the response time    depends on the remaining wait time for the prefetching request's return. However, there were only 5 such ``wait'' requests among 425 total requests in the 32 apps. This strongly suggests that \appr's choice for the placement of Trigger Points is effective in practice.

\begin{table}[t]
  \caption{Results of \appr's evaluation across the 32 third-party apps.}
\vspace{-3mm}
  	\label{tbl:real_app}
  	\centering
 	\resizebox{.99\linewidth}{!}{
\begin{tabular}{|l|c|c|c|c|}
\hline
                 & Min.     & Max.     & Avg.     & Std. Dev. \\ \hline
Runtime Requests & 1       & 64      & 13.28   & 14.41     \\ \hline
Hit Rate         & 7.7\%   & 100\%   & 47.76\% & 28.81\%   \\ \hline
Latency Reduction        & 87.41\% & 99.97\% & 98.82\% & 2.3\%     \\ \hline
\end{tabular}
 	}
\end{table}
\vspace{-1mm}
\section{related work}
\label{related_work}
\vspace{-1mm}

Prefetching of HTTP requests has been applied successfully in the browser domain~\cite{wang2016speeding, mickens2010crom, netravali2016polaris, wang2013mobile}. Unfortunately, approaches targeting page load times cannot be applied to mobile apps. The bottleneck for page load times is resource loading~\cite{wang2011web}, because one initial HTTP request will require a large number of subresources (e.g., images), which can only be discovered after the main source is fetched and parsed. Thus, existing research efforts have focused on issues such as prefetching subresources~\cite{wang2012far,http2push},  developer support for speculative execution~\cite{mickens2010crom}, and restructuring the page load process~\cite{wang2016speeding}. In mobile apps, the HTTP requests are always light-weight~\cite{li2016automated}: one request  only fetches a single resource 
that does not require any further subresource fetching. Therefore, our work focuses on prefetching the future requests that a user may trigger  rather than the subresources within a single request.

Researchers have recently begun exploring prefetching in the mobile app domain. One research thread has attempted to answer ``how much'' to prefetch under different contexts (e.g., network conditions)~\cite{yang2017prefetch,higgins2012informed,Baumann2017}, while assuming that ``what'' to prefetch is handled by the apps already. Another thread of work focuses on fast prelaunching by trying to predict what app the user will use next~\cite{Parate2013,Yan2012fastlaunch,Baeza-Yates2015}. By contrast, our work aims to provide an automated solution to determine ``what'' and ``when'' to prefetch for a given app in a general case. As discussed previously, other comparable solutions---server-based~\cite{http2push,de2007improving,padmanabhan1996using}, human-based~\cite{mickens2010crom,li2014reflection}, history-based~\cite{padmanabhan1996using,behaviorcharacterization,fan1999web,Kostakos2016,WhiteSearchResultPrefetch}, and domain-based~\cite{earlybird,spice:socialdrivenprefetching,Battle2016}---have limitations which we directly target in \appr. 

To the best of our knowledge, \appr is the first technique to apply program analysis to  prefetching HTTP requests in mobile apps in order to reduce user-perceived latency. Bouquet~\cite{li2016automated} has applied program analysis techniques to bundle HTTP requests in order to reduce energy consumption in mobile apps. Bouquet detects \textit{Sequential HTTP Requests Sessions} (SHRS), in which the generation of the first
request implies that the following requests will also be made, and then bundles  the requests together to save energy. This can be considered a form of prefetching. However, this work does not address inter-callback analysis and the SHRS are always in the same callback. Therefore, the ``prefetching'' only happens a few statements ahead (within milliseconds most of the time) and has no tangible effect on app execution time. 

\vspace{-1mm}
\section{conclusion and future work}
\label{conclusion}
\vspace{-1mm}

We have presented \appr, a novel program analysis-based technique that reduces the user-perceived latency in mobile apps by prefetching certain HTTP requests. While \appr cannot be applied to all HTTP requests an app makes at runtime, it provides significant performance savings in practice. Several of \appr's current facets make it well suited for future work in this area, both by us and by others. For instance, \appr defines formally the conditions under which the requests are prefetchable. This can lead to guidelines that developers could apply to make their apps more amenable to prefetching, and lay the foundations for further program analysis-based prefetching techniques. We have also identified several shortcomings to \appr whose remedy must include improvements to string analysis and callback analysis techniques. Another interesting direction is to improve the precision and reduce the waste associated with prefetching by incorporating certain dynamic information (e.g., user behavior patterns, runtime QoS conditions). Finally, \appr's microbenchmark (MBM) forms a foundation for standardized empirical evaluation and comparison of future efforts in this area. 
\vspace{-1mm}
\section{acknowledgment}
\vspace{-1mm}

We would like to thank William G.J. Halfond, Atanas Rountev, Yuhao Zhu, and their research groups. This work is supported by the U.S. National Science Foundation under grants no. CCF-1618231 and CCF-1717963, U.S. Office of Naval Research under grant no. N00014-17-1-2896, and by Huawei Technologies Co., Ltd.

\clearpage
\bibliographystyle{ACM-Reference-Format}
\bibliography{icse2018ref} 

\end{document}